\documentclass[showpacs,preprintnumbers,amsmath,amssymb,prb]{revtex4}

\usepackage{graphicx}
\usepackage{dcolumn}
\usepackage{bm}

\begin{document}

\title{ Partially disordered state near ferromagnetic transition in $Mn Si$}

\author{S.V.Maleyev}
 \email{maleyev@sm8283.spb.edu}
\author{S. V. Grigoriev}
\affiliation{Petersburg Nuclear Physics Institute, Gatchina, Leningrad District 188300, Russia}

\date{\today}

\newcommand{\m}{\mathbf}
\begin{abstract}

The polarized neutron scattering in helimagnetic MnSi at low $T$ reveals existence of a partially disordered chiral state at ambient pressure in the magnetic field applied along $\langle 111\rangle$ axis below the first order transition to the non-chiral ferromagnetic state. This unexpected phenomenon is explained by the analysis of the spin-wave spectrum. We demonstrate that the square of the spin-wave gap becomes negative under magnetic field applied along $\langle 111\rangle$ and $\langle 110\rangle$ but not along the $\langle 100\rangle$ direction. It is a result of competition between the spin-wave interaction and cubic anisotropy. This negative sign means an instability of the spin wave spectrum for the helix and leads to a destruction of the helical order, giving rise to the partially disordered state below the first order ferromagnetic transition.

\end{abstract}  

\maketitle
Non-centrosymmetric cubic helimagnets such as $Mn Si, Fe Ge, Fe Co Si$ are the subject of the intensive experimental and theoretical studies for the last several decades. Their single-handed helical structure was explained by Dzyaloshinskii \cite{D}. The full set of interactions responsible for observed helical structure (Bak-Jensen model) was established later in \cite{N,B} in agreement with existing experimental data (see for example \cite{I} and references therein). The renascence in this field began with a discovery of the quantum phase transition to a disordered (partially ordered) state in $Mn Si$  at high pressure \cite{P1} and \cite{K}. The following properties of this state attract the main attention: i) non-Fermi-liquid conductivity, ii) spherical neutron scattering surface with the weak maxima along the $\langle 1 10\rangle$ axes \cite{P2}, \cite{P3}, whereas at ambient pressure Bragg reflections are observed along  $\langle 111 \rangle$ \cite{I}. These features and the structure of the partially ordered state were discussed in several theoretical papers (see \cite{T,R,BK,Bn} and references therein). It should be noted also that the spherical scattering surface with maxima along $\langle 111 \rangle$ was observed at ambient pressure just above critical temperature $T_c\simeq 29K$. This experiment was explained using the Bak-Jensen model \cite{G1}.

These studies shadowed an important problem of the helix structure evolution in the external magnetic field $H$. In particular, simple phenomenological \cite{Pl} and microscopical \cite{M} theories  predict the smooth second order transition from the conical to the ferromagnetic state. The spin component of the cone parallel to the applied field is proportional to the magnetization and it increases as ($H/H_{C}$) up to its saturated value. This prediction is in   agreement with experiment \cite{K}. The perpendicular, rotating spin components fade away with the field and according to the plain theory the helical Bragg reflections must decrease as $(H_C^2-H^2)/H_C^2$ where $H_C$ is the critical field for the ferromagnetic transition.  This, however, contradicts to the experimental facts if $\m H\parallel [111]$ (See \cite{G2} and Fig.1). The experiment shows that the transition is of the first order and the Bragg intensity does not follow the law given above.  

In this paper we demonstrate that although from a general expression for the ground state energy one can expect the second order transition at the critical field $H_C$, this is true for $\m H \parallel [100]$ only but the situation changes if $\m H \parallel [111]$ or $\m H \parallel [110]$. In the last cases the spin-wave spectrum is unstable in the field range $H_1<H<H_C$ due to the cubic anisotropy as the square of the spin-wave gap becomes negative and the helical long-range order decays in this field interval. The ferromagnetic state occurs above $H_C$. Hence we have a region below $H_C$ where the partially disordered state coexists with the almost saturated magnetization.  It is a reason why this state has not been noticed in the earlier macroscopic measurements \cite{P1}, \cite{K}.

Let us consider the conical helix with the lattice spin   
$\m{S_R}=S^\zeta_\m R\hat \zeta_\m R+S^\eta_\m R\hat \eta_\m R+S^\xi_\m R\hat \xi_\m R$ where
\begin{equation}
\begin{aligned}
\hat\zeta_\m R&=\hat c\sin\alpha+(\m A e^{i\m{k\cdot R}}+c.h.)\cos\alpha;\\
\hat \eta_\m R &=i(\m A e^{i\m{k\cdot R}}-c.h.);\\
\hat \xi_\m R&=\hat c\cos\alpha-(\m A e^{i\m{k\cdot R}}+c.h.)\sin\alpha,
\end{aligned}
\end{equation}
where $\m A=(\hat a-i\hat b)/2$ and the unit vectors $\hat \zeta,\hat \eta$ and $\hat \xi$ form the right-handed frame. If $\alpha=0$ we have a plain helix \cite{M}. The spin operators are given by the well known expressions: $S^\zeta_\m R=S-(a^+a)_\m R$, $S^\eta_\m R=-i(S/2)^{1/2}[a_\m R-a^+_\m R-(a^+a^2)_\m R/(2S)]$ and $S^\xi_\m R=(S/2)^{1/2}[a_\m R+a^+_\m R-(a^+a^2)_\m R/(2S)]$.

Similar to in \cite{M}, we use the following Hamiltonian
\begin{equation}
\begin{aligned}
H&=\frac{1}{2}\sum\{-J_\m q \m S_\m q\cdot \m S_\m{-q}+2iD_\m q [\m S_\m q\times\m S_\m{-q}]\\&
+F_\m q\sum_{l=x,y,z} S_{\m q,l}S_{-\m q,l}q^2_l\}+K\sum_{l=x,y,z} S_{\m q,l}S_{\m p,l}S_{\m h,l}S_{\m f,l}\\&+N^{1/2}\m H\cdot\m S_0,
\end{aligned}
\end{equation}
where the first term is the ferromagnetic exchange interaction, the second term is the Dzyaloshinskii interaction (DI) responsible for the helix structure \cite{D}. The following terms are the anisotropic exchange, the cubic anisotropy and the Zeeman energy. They determine the orientation of the helix vector $\m k$ in the magnetic field (see \cite{M}, \cite{G2} and \cite{G3}). The following hierarchy of interactions holds
\cite{N,B}: $J_0>>D_0/a>>F_0/a^2\sim K$ where $a$ is the lattice constant. Replacing $\m S_\m R\to S\hat \zeta_\m R$ one gets the classical energy \cite{M}
\begin{equation}
\begin{aligned}
E_{cl}&=S\left\{\frac{A k^2}{2}+\frac{S F_0 k^2 L(\m k)}{2}\right. \\&\left.-S D_0(\m k\cdot [\hat a\times\hat b])\right\}cos^2\alpha+S H_\parallel\sin\alpha+E_{cub}
\end{aligned}
\end{equation}
where $A=S(J_0-J_\m k)/k^2$ is the spin-wave stiffness at  $q>>k$, $L(\m k)=\sum \hat k^2_l(\hat a^2_l+\hat b^2_l)$ and $H_\parallel$ is the field component parallel to $\m k$. 

Using  (3) and (4) and taking into account that the single ion contribution $E_{cub}$ does not depend on $\m k$ we get \cite{Er2}
\begin{equation}
	\begin{aligned}
	k_l&=k(D_0/|D_0|)[\hat c_l-S F_0(\hat a^2_l+\hat b^2_l)/(2A)]\\
	E_{cl}&=-(S A k^2/2)[1-S F_0L(\hat c)/(2A)]\cos^2\alpha+H_\parallel\sin\alpha+E_{cub},
	\end{aligned}
\end{equation}
 where $k=S|D_0/A|$ and $L(\hat c)$ is a cubic invariant. For $D_0>0$ or $D_0<0$ we have the right or the left helix, respectively \cite{B}. For $F_0<(>)0$ the helix vector $\m k$ is oriented along the $\langle 111\rangle \quad (\langle 100 \rangle )$ axes as the invariant $L$ has two extrema $2/3(0)$ \cite{B}.
 
Neglecting $E_{cub}$ in Eq.(4) one obtains the conical state with $\sin\alpha=-H_\parallel/H_C$ if $H_\parallel<H_C$ where  $H_C=A k^2(1-S F_0L/(2A)]\simeq A k^2$ and the ferromagnetic state for $H_\parallel>H_C$ \cite{M,Dip}. It can be shown that $E_{cub}$ gives a negligible contribution to $H_C$ and to $\sin\alpha$. The principal parameters of the magnetic structure for MnSi are: $A\simeq 52meV\mbox{\AA}^2$, $k\simeq 0.038\mbox{\AA}$ and $H_C\simeq 0.6T\simeq A k^2$ in agreement with the theory (see\cite{M}). Another energy  $F_0 k^2\sim 0.01meV\simeq 0.1T$ was estimated from the anisotropy of the critical neutron scattering \cite{G1} and from the reorientation of the helix axis in the magnetic field \cite{G2}. The value of $K$ will be estimated below.

The classical energy depends on $H_\parallel$ only. The general expression for the ground state energy, derived in \cite{M}. It contains yet another term of purely quantum nature, which depends on the spin-wave gap $\Delta$ as a parameter. As shown in \cite{M} if $H_{\perp} > \Delta\sqrt{2}$ the helix wave vector $\m k$ is directed along the magnetic field. For MnSi $\Delta\simeq 12\mu eV \simeq 0.1T \ll H_C$  \cite{G2}. 

In a linear spin-wave theory the gap appears due to the cubic anisotropy but it equals to zero at $H_\parallel=0$ \cite{Er1}. There is yet another contribution to the gap, which is a result of the spin-wave interaction. We begin with the former. For evaluation of $\Delta^2$ one has to consider the uniform part of the bilinear Hamiltonian, which is given by
\begin{equation}
	H_0=E_0a^+_0a_0+B_0(a_0^2+a_0^{+2})/2,
\end{equation}
and  $\Delta^2_0=E_0^2-B_0^2$. If one neglects the cubic anisotropy at $H_\parallel<H_C$ then one has $E_0=B_0=(H_C/2)\cos^2\alpha$ and the gap is zero \cite{M}, \cite{Er2}. Taking into account $H_{cub}$, one obtains after simple but rather tedious calculations
\begin{equation}
\begin{aligned}
E_0&=(H_C\cos^2\alpha)/2+E_1+E_2; B_0=(H_C\cos^2\alpha)/2-E_2\\
E_1&=-4\Lambda[(1-L)\sin^4\alpha+3L\sin^2\alpha \cos^2\alpha\\
	&+(3/8)(2-L)\cos^4\alpha];\\ E_2&=(3\Lambda/4)[4L\sin^2\alpha+(2-L)\cos^2\alpha];\\
\Delta^2_{Cub}&=(H_C\cos^2\alpha+E_1)(E_1+2E_2),
\end{aligned}
\end{equation}
where $\Lambda=S^3K$. These expressions hold at $H_\parallel<H_C$ and, indeed, 
$\Delta^2_{Cub}=0$ in zero field only \cite{Er1}.
 
For $H\ge H_C$ we have $\cos\alpha=0$ and if without cubic anisotropy $E_0=H-H_C,\quad B_0=0$ and the gap $\Delta=H-H_C$ \cite{M}, while if taking it into account the cubic anisotropy we get 
\begin{equation}
\Delta^2_{Cub}=[H-H_C-4\Lambda(1-L)][H-H_C+\Lambda(10L-4)].
\end{equation}
For the two principal directions $L_{111}=2/3$ and $L_{001}=0$ and we obtain:
\begin{equation}
\begin{aligned}
	\Delta^2_{Cub,[1,1,1]}&=(H-H_C-4\Lambda/3)(H-H_C+8\Lambda/3);\\
	\Delta^2_{Cub,[1,0,0]}&=(H-H_C-4\Lambda)^2,
	\end{aligned}
\end{equation}
Thus, one comes to the important conclusion: $\Delta^2_{Cub,[1,1,1]}$ is negative at $H=H_C$ for both signs of $K$ \cite{Com1}. This circumstance is decisive for the stability of the system if one takes into account that the contribution to $\Delta^2$, stem from the spin-wave interaction, is proportional to $\cos^4\alpha$ and disappears at $H$ close to $H_C$.

Let's consider the spin wave interaction to the gap $\Delta^2_{Int}$. As the DI breaks the total spin conservation law it must lead to the spin-wave gap. However, in cubic crystals this interaction is very soft [see Eq.(2)] and the gap appears as a result of the spin-wave interaction only similar to the case of pseudo-dipolar interaction in antiferromagnets \cite{Pt}, \cite{Com}. In the one-loop approximation it consists of both the Hartree-Fock (HF) part evaluated in \cite{M} at $H=0$ and the second order contribution from three-point spin-wave interaction, which appears due to the helical structure. It was ignored in \cite{M}.  The diagrams for both contributions are shown in Fig.2  where lines correspond to Green functions $G_\m q=-<Ta_\m q,a^+_\m q>$ and $F_\m q=F^+_\m q=-<Ta_\m q,a_{-\m q}>$, which in $\omega$-representation  are given by \cite{M}:
\begin{equation}
	G_\m q(i\omega)=\frac{E_\m q+i\omega}{(i\omega)^2-\epsilon^2_\m q};\quad F_\m q(i\omega)=-\frac{B_\m q}{(i\omega)^2-\epsilon^2_\m q},
\end{equation}
	where $E_\m q=S(M_o-M_\m q)+B_\m q; B_\m q=(S/2)(M_\m q -J_\m q)\cos^2\alpha; M_\m q=J^{+}_\m q+2D_\m q(\m k\cdot\hat c)$,  $J^{\pm}_\m q=(J_\m{q+k}\pm J_\m{q-k})/2$ and the spin-wave energy $\epsilon_\m q=(E^2_\m q-B^2_\m q)^{1/2}$. Although at  $q\ll 1/a$ these expressions give the same result as obtained in \cite{M}: $E_\m q=A q^2+B_\m q;\quad B_\m q =(A k^2/2)\cos^2\alpha$  and $\epsilon_\m q=A q(k^2\cos^2\alpha+q^2)^{1/2}$. However for the present consideration we need them for all $\m q$ as the formulae for $\Delta^2_{Int}$ contains the sums, which saturate at $q\sim 1/a$.
	
The contribution of the forth-point interaction to $\Delta^2$ was analyzed in \cite{M} at $H=0$ and $T=0$. At an arbitrary $H$ the gap consists of two terms 
	\begin{equation}
	\begin{aligned}
	V_1&=(1/4N)\sum(M_\m 1+M_2-M_\m{1-3}-M_\m{2-3})a^+_\m 1a^+_\m 2a_\m 3 a_\m 4;\\
	V_2&=(1/4N)\sum(M_\m 1-J_\m 1)[2(a^+a)_\m 1(a^+a)_\m{-1}\sin^2\alpha\\
	&-(a^+a^2)_\m 1(a_\m 1+a^+_\m {-1})\cos^2\alpha],
	\end{aligned}
\end{equation}
where $\m 1=\m q_1$ etc. At small momenta we have in parentheses $-2A\m{(3\cdot 4)}/S$ and $A k^2/S$ for $V_1$ and $V_2$, respectively.

Now one has to consider the $V_1$ interaction and the second part of $V_2$ interaction together. They give the principal contribution to the gap $\Delta^2_{Int}$ 
\begin{equation}
	\Delta^2_{Int}=\frac{S A k^2\cos^4\alpha}{8N}\sum (M_\m q-J_\m q)=\frac{(A k^2)^2\cos^4\alpha}{4S N}\sum\frac{D_\m q}{D_0},
\end{equation}
where in r.h.s we take into account that $\sum J^+_\m q=\sum J_\m q=0$ and according to Eq.(4) $(\m k\cdot\hat c)=A k^2/(S D_0)$.  This contribution is $T$-independent.

The contribution of the first term in $V_2$ is more complicated one. Its $T$-independent part is proportional to $(k a)^2$ and may be neglected [for $Mn Si$  $(k a)^2\approx 0.03$]. The $T$-dependent contribution consists of two parts. The excitations with $q\gg k$ have a quadratic dispersion $\epsilon_\m q\approx A q^2$ \cite{M} and at $T\gg A k^2$ they are responsible for the first part, which has the form $\Delta^2_{2,1} =(1/2)(A k^2)^2\zeta(3/2)(k a)^3[T/(2\pi A k^2)]^{3/2}\sin^2\alpha\cos^2\alpha$. In spite of small factor $(k a)^3$ it may be important at sufficiently high $T$.

The second part is not so trivial. According to \cite{M} (cf. also \cite{BK}) at   $q\lesssim k$ and $H<H_C$ the spin-wave spectrum becomes strongly anisotropic due to umklapp processes connecting excitations with $\m q$ and $\m{q\pm k}$. In zero field taking into account the gap we have 
\begin{equation}
\epsilon_\m q=[A^2(k^2q^2_\parallel+3q^4_\perp/8)+\Delta^2]^{1/2}.
\end{equation}
As the field increases this anisotropy becomes weaker since the term in the expression for $\epsilon_\m q$ appears to be proportional to $q^2_\perp$. In the ferromagnetic state ($H>H_C$) this anisotropy vanishes. In a very weak field we can use Eq.(12) and then the second part is given by 
\begin{equation}
\Delta^2_{2,2}=(1/\pi)A k^2 T\sqrt{3/2}\ln(A k^2/\Delta)\sin^2\alpha\cos^4\alpha.	
\end{equation}
This expression diverges if $\Delta\to 0$. Similar divergences were discussed in \cite{KB}. However, we can neglect this contribution due to the small factor $(k a)^3$ and decreasing of logarithm when $H$ increases.

 The three-point interaction is given by  $V_3=V_-\cos\alpha+V_+\sin\alpha\cos\alpha$ where $V_{\pm}=(2S/N)^{1/2}\sum C_\m q^{(\pm)}(a^+a)_\m{-q}(a_\m{- q}\pm a^+_\m q)$ and
\begin{equation}
\begin{aligned}
C^{(-)}_\m q&=[J^{(-)}_\m q+D_\m q(\m q\cdot\hat c)]\sim (A/S)\m{(q\cdot k)}(q a)^2\\
C^{(+)}_\m q&=[(J_\m q-J_\m q^{(+)})/2-D_\m q(\m k\cdot \hat c)]\simeq-A k^2/(2S),	
\end{aligned}
\end{equation}
where  the r.h.s. expressions are derived from Eq.(4) at $q\ll 1/a$. The corresponding contributions to $\Delta^2$ were evaluated as in \cite{Pt}.  The $T$-independent term is proportional to $(k a)^2$ and the second one is equal to $-2\Delta_{2,2}/3$. Both of them are small and can be neglected. Finally we have
\begin{equation}
	\Delta^2=\Delta^2_{Int}+\Delta^2_{Cub}.
	\end{equation}
		
The field dependence of the ratio $\Delta^2(H)/\Delta^2(0)$ for the three directions $\langle111\rangle$, $\langle100\rangle$ and $\langle 110\rangle$ are shown in Fig.3 with  $H_{C}=0.565T,\quad \Delta(0)=0.1$ T in agreement with the experiment. The cubic anisotropy $\Lambda$ was chosen to be equal to $-0.05$ T. In case of $ \m H \parallel [100]$ $\Delta^2$ remains positive and the spin-wave spectrum is stable at all $H$. In this case at $H>H_{C}$ along with ferromagnetic spin configuration the spin-wave components of the lattice spins has to remain rotating [see Eq.(1)] as was discussed in \cite{M}. For $\langle111\rangle$ direction $\Delta^2$ is negative at $H>H_1\simeq 0.72 H_{C}$ and the spin-wave spectrum becomes unstable. Hence the long-range helical order demolishes and the corresponding Bragg peaks disappear and the scattering has to be spread around them.  Such decrease of the intensity along $ [111]$ is shown in Fig.1 in qualitative agreement with the theory. However, more detailed measurements are needed. A similar instability has to be along $ [110] $ also contrary to the expectations for ${\bf H} \parallel [100] $.

The width of this scattering may be estimated from the condition $\epsilon^2_\m q=0$, as for larger $q$ the spin-wave spectrum can not feel the disorder. Near $H_1$ we have $\epsilon^2_\m q\sim (A q k)^2+\Delta^2$ and the inverse correlation length of the disorder $\kappa=k(|\Delta|/A k^2)\ll k$. Close to $H_{C}$ $\epsilon_\m q\sim A q^2$ \cite{M} and $\kappa=k(|\Delta|/A k^2)^{1/2}$.

This disordered state have a strong chirality, which is demonstrated by a constant polarization of the scattered neutrons in the whole field range below $H_{C}$ (see inset in Fig.1). This polarization is determined as $P=(I_{-}-I_{+})/(I_{-} +I_{+})$ and according to the general theory \cite{M1} the ratio $P/P_0$, where  $P_0$ is the initial neutron polarization, is  the chirality at given $\m q$. A strong drop of the polarization at $H_{C}$ is a signature of the first order transition to the uniform ferromagnetic state with weak chiral fluctuations.

In conclusion, we analyzed thoroughly the field behavior of the spin wave gap in the spin-wave spectrum of cubic helimagnets. It is shown that if the field is applied parallel to both $ [111]$ and $[ 110]$  directions a partially disordered state has to take place at $H_1<H_{C}$.  We demonstrated that this state appears when the square of the spin-wave gap becomes negative and the spin-wave spectrum unstable. We presented the first observation in MnSi of this partially disordered chiral state in the magnetic field.
	
	The work is supported in part by the RFBR (projects No 05-02-19889, 06-02-16702 and 07-02-01318) and the Russian State Programs "Quantum Macrophysics" and "Strongly correlated electrons in Semiconductors, Metals, Superconductors and magnetic Materials" and Russian State Program "Neutron research of solids".

\begin{figure}
\centering
\includegraphics[scale=0.4]{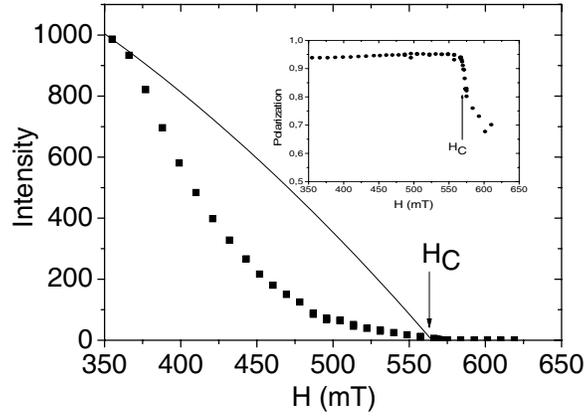}
\caption{The intensity of the Bragg reflection in $Mn Si$ at $T=15 $ K as function of the field at $ \m H \parallel [111]$. The full line is the theoretical prediction (see the text). Inset: the spin chirality as a function of the field measured by polarized neutrons(see text).} 
\end{figure}

\begin{figure}
\centering
\includegraphics[scale=0.4]{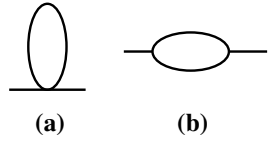}
\caption{Hartree-Fock (a) and three-point diagrams for the spin-wave gap (b).} 
\end{figure}

\begin{figure}
\centering
\includegraphics[scale=0.4]{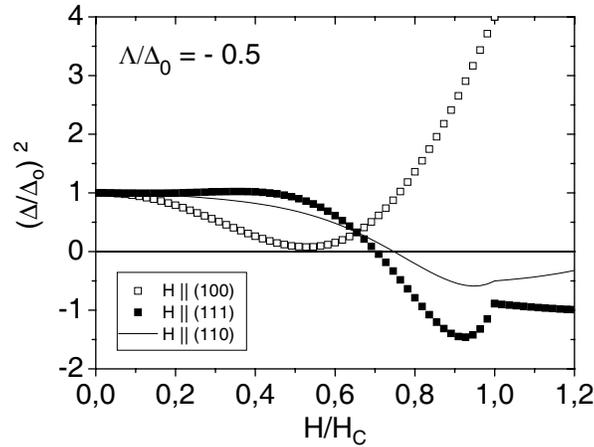}
\caption{The magnetic field dependence of the ratio $\Delta^2(H)/\Delta^2(0)$. Parameters are given in the text.} 
\end{figure}


\begin{thebibliography}{99}
\bibitem{Er2} There are misprints in Eqs. (13) and (21,22) in Ref. \cite{M}: the $F$ terms has to have negative sign.
\bibitem{Dip} Due to demagnetization $H_C$ has additional  term depending on the sample form \cite{M}.
\bibitem{Com} Results presented below were obtained  as in \cite{Pt} using  modification of the Belyaev's
technique \cite{AA} adjusted to non-Hermitian spin-wave interaction in Dayson-Maleyev representation.
\end{thebibliography}
\end{document}